\documentclass[10pt, a4paper, twocolumn]{article} 

\usepackage[english]{babel} 

\usepackage{microtype} 

\usepackage{amsmath,amsfonts,amsthm} 

\usepackage[svgnames]{xcolor} 

\usepackage[]{caption} 

\usepackage{booktabs} 

\usepackage{lastpage} 

\usepackage{graphicx} 

\usepackage{enumitem} 
\setlist{noitemsep} 

\usepackage{sectsty} 
\allsectionsfont{\usefont{OT1}{phv}{b}{n}} 

\usepackage{geometry} 

\usepackage[T1]{fontenc} 
\usepackage[utf8]{inputenc} 

\usepackage{XCharter} 

\usepackage{fancyhdr} 
\fancypagestyle{firstpage}{ 
	\fancyhf{}
}

\newcommand{\authorstyle}[1]{{\large\usefont{OT1}{phv}{b}{n}\color{DarkRed}#1}} 

\newcommand{\institution}[1]{{\footnotesize\usefont{OT1}{phv}{m}{sl}\color{Black}#1}} 

\usepackage{titling} 

\newcommand{\HorRule}{\color{DarkGoldenrod}\rule{\linewidth}{1pt}} 

\pretitle{
	\vspace{-30pt} 
	\HorRule\vspace{10pt} 
	\fontsize{28}{15}\usefont{OT1}{phv}{b}{n}\selectfont 
	\color{DarkRed} 
}

\posttitle{\par\vskip 15pt} 

\preauthor{} 

\postauthor{ 
	\vspace{10pt} 
	\par\HorRule 
	\vspace{-20pt} 
}

\usepackage{lettrine} 
\usepackage{fix-cm}	

\newcommand{\initial}[1]{ 
	\lettrine[lines=3,findent=4pt,nindent=0pt]{
		\color{DarkGoldenrod}
		{#1}
	}{}%
}

\usepackage{xstring} 

\newcommand{\lettrineabstract}[1]{
	\StrLeft{#1}{1}[\firstletter] 
	\initial{\firstletter}\textbf{\StrGobbleLeft{#1}{1}} 
}

\title{Terahertz whispering gallery mode bubble resonator} 

\author{
	\authorstyle{Dominik Walter Vogt\textsuperscript{1,*} and Rainer Leonhardt\textsuperscript{1}} 
	\newline\newline 
	\textsuperscript{1}\institution{Dodd-Walls Centre for Photonic and Quantum Technologies, Department of Physics, The University of Auckland, Private Bag 92019, Auckland 1142, New Zealand}\\
	\textsuperscript{*}\institution{Corresponding author: d.vogt@auckland.ac.nz}
} 

\date{} 

\begin{document}

\maketitle 

\thispagestyle{firstpage} 


\lettrineabstract{Whispering gallery mode (WGM) resonators are compelling optical devices, however they are nearly unexplored in the terahertz (THz) domain. In this letter, we report on THz WGMs in quartz glass bubble resonators with sub-wavelength wall thickness. An unprecedented study of both the amplitude and phase of THz WGMs is presented. The coherent THz frequency domain measurements are in excellent agreement with a simple analytical model and results from numerical simulations. A high finesse of 9 and a quality (Q) factor exceeding 440 at 0.47\,THz are observed. Due to the large evanescent field the high Q-factor THz WGM bubble resonators can be used as a compact, highly sensitive sensor in the intriguing THz frequency range.}



Whispering gallery mode micro ring resonators and WGM micro bubble resonators are of enormous importance in the optical regime. They show unprecedented high Q-factors and provide excellent opportunities for integrated photonic systems. Typical applications include, but are not limited to, spectral switches and filters, optical delay lines, lasers and sensors \cite{silicon_microring, matsko2006optical, yang2014quasi}. However, little work has yet been done in the THz realm on WGM resonators \cite{Zhang:02,Zhang:03,Preu:08,Preu:13}, although they provide significant advantages compared to their equivalents in the optical regime. The strong requirements on high precision manufacturing and experimental implementation (e.g. nm-positioning and thermal insulation) are highly alleviated due to the large wavelength in the THz frequency range. Moreover, the coherent detection of THz radiation 
allows easy access to both amplitude and phase of the spectral properties of the WGM resonators and therefore providing more information about the system. But THz WGM resonators also advance into a novel regime in which only a few wavelengths are confined in the circumference of the resonators. In consequence, effects like bend loss and waveguide dispersion become more important.  

In this letter, we present the observation of THz WGMs in a dielectric bubble resonator with sub-wavelength wall thickness. The thin-walled spherical resonator design limits the impact of the high material loss, essential for the realization of high Q-factor resonators in the THz frequency range. Evanescent coupling into the weakly confined WGMs can be achieved using a single mode sub-wavelength dielectric fiber. Finally, the design provides a potential platform to exploit the sensing properties of bubble resonators utilizing the unique spectral properties of many gases and liquids in the THz frequency domain. 

The dielectric WGM THz bubble resonators are made of quartz glass with a low OH-content (<\,30\,ppm). A low material loss of about 0.2\,${\textrm{cm}}^{-1}$ at 0.45\,THz and a high refractive index of ${n}_{\textrm{quartz}}$ = 1.96 (0.1 - 1\,THz) render quartz glass an excellent choice for the realization of the THz bubble resonators \cite{material}. From a practical point of view, quartz glass conveniently allows the production of mechanically stable bubble resonators by means of standard glass blowing techniques. By this fabrication method, homogeneously shaped THz bubble resonators can be easily achieved due to the relatively large wavelength of THz radiation. Commercially available quartz glass capillaries with a diameter of 3.7\,mm and a wall thickness of 0.14\,mm are used as preforms for the bubble resonators.
\begin{figure}[ht]
\centering
\fbox{\includegraphics[width=\linewidth]{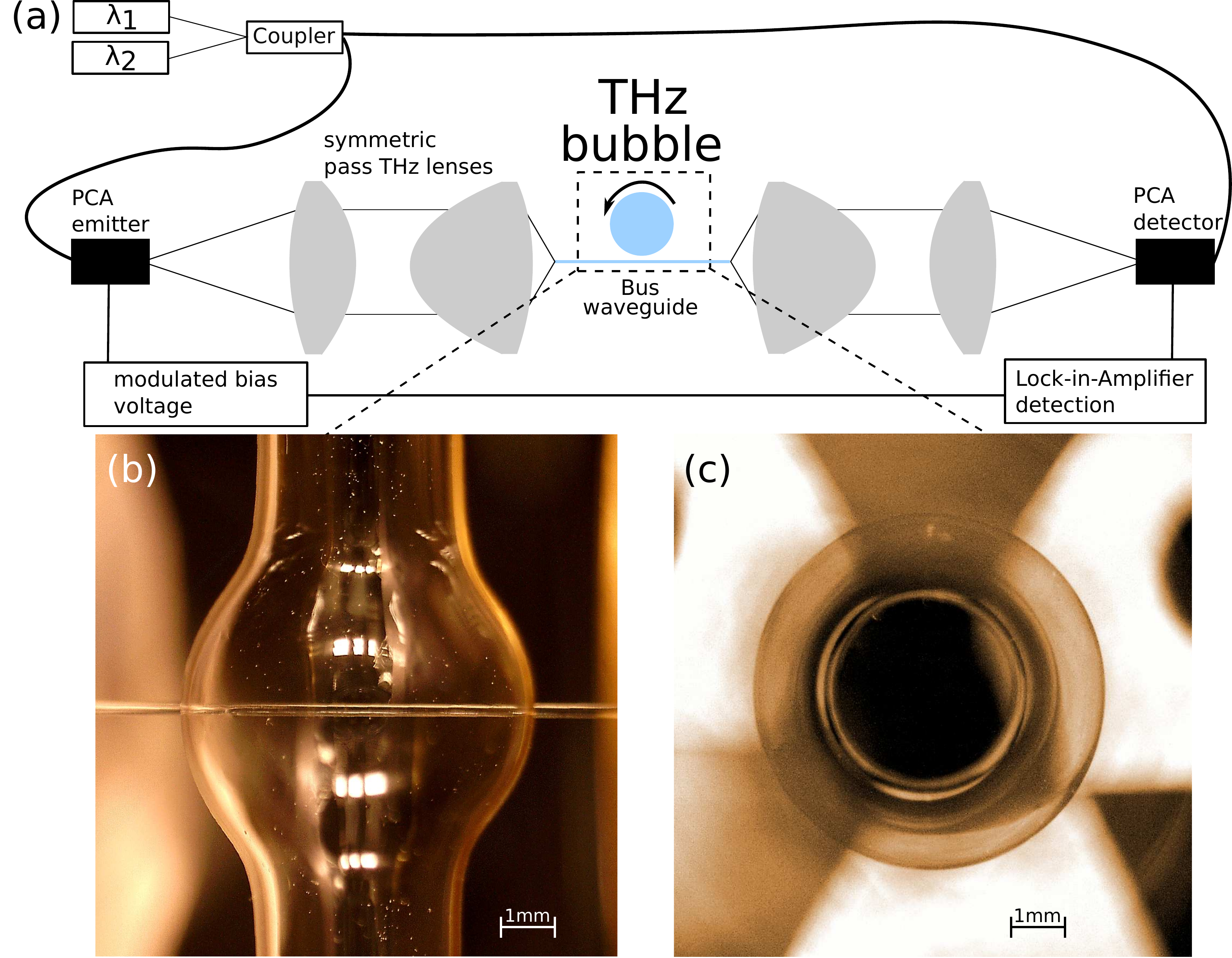}}
\caption{(a) Schematic diagram of the experimental setup for the spectral characterization of the dielectric WGM THz bubble resonator. The WGMs are excited via evanescent coupling from a bare 200\,µm optical fiber. Symmetric pass lenses are deployed to couple the THz radiation into the sub-wavelength fiber and (b), (c) show a microscope image of a THz bubble resonator from side view and top view, respectively.}
\label{fig:1}
\end{figure}
The bubble resonator's wall thickness is characterized using the 1.3\,µm source of an optical coherence tomography setup \cite{Brauer:16}. A THz bubble resonator is shown in the inset of fig. \ref{fig:1} with an outside diameter of 6.3\,±\,0.1\,mm and a sub-wavelength wall thickness of 77\,±\,2\,µm ($\sim\,{\lambda}_{\textrm{quartz}}/4$ at 0.5\,THz). The nearly perfect spherical shape in the equatorial region is easily recognizable. 

The quartz glass THz bubble resonators are experimentally characterized with a tunable continuous wave (CW) THz system based on optical heterodyning (Toptica TeraScan 1550nm). An absolute frequency resolution of <\,2\,GHz combined with a relative frequency resolution of <\,10\,MHz ensure a sufficient resolution for the study of the high Q-factor THz WGMs. A schematic diagram of the experimental setup is shown in fig. \ref{fig:1} (a). Coherent generation and detection of linear polarized THz radiation is done using standard fiber coupled photoconductive antennas (PCAs). The THz radiation is collimated and focused with short focal length (25\,mm) symmetric-pass THz lenses to efficiently couple the THz radiation into a bus waveguide \cite{lens}. It consists of a single mode silica optical fiber with a diameter of 200\,µm as shown in fig. \ref{fig:1} (b). The WGMs of the THz bubble are excited via evanescent coupling from the sub-wavelength fiber. To enhance the overlap of the evanescent fields, the polymer 
cladding of the fiber is removed. A 2D computer controlled translation stage allows one to precisely position the THz bubble resonator in respect to the bus waveguide. The position of the bubble resonator is monitored using two microscope cameras (top and sideview). The polarization of the THz radiation is out of the plane of fig. \ref{fig:1} (a) and (c) (in the plane of fig. \ref{fig:1} (b)) which ensures an excitation of the $\textrm{TE}$ modes (electric field tangential to the bubble wall) of the bubble resonator. 

To characterize the resonator, the transmitted signals of the bus waveguide with and without resonator in close proximity to the bus waveguide are measured. The coherent detection with the THz frequency-domain system (FDS) allows one to study the intensity and the relative phase of the calculated transmission ratio of the bus waveguide. The intensity ratio is fitted with the well-known analytical expression for the transmission function of a ring resonator $T$  \cite{silicon_microring}. T is the function describing the ratio of the transmitted intensities with and without coupling to the resonator, respectively:

\begin{equation}
T = \frac{a^2-2ar\cos{\phi}+ r^2}{1-2ar\cos{\phi} + (ra)^2}
\label{eq:amplitude}
\end{equation}

with $a$ the single-pass amplitude transmission and $r$ the self-coupling coefficient. $\phi = \beta L$ is the phase shift for one roundtrip, $\beta$ is the propagation constant and $L$ the roundtrip length. Please note that a ring resonator with no intrinsic losses is modeled with $a=1$.

Furthermore, the measured phase shift is compared to the effective phase shift $\varphi$ obtained from the analytical model \cite{silicon_microring}:
\begin{equation}
\varphi= \pi + \phi + \arctan{\frac{r\sin{\phi}}{a-r\cos{\phi}}}+\arctan{\frac{ra\sin{\phi}}{1-ra\cos{\phi}}}
\label{eq:phase}
\end{equation}

While the analytical model for a ring resonator fails to predict higher order azimuthal modes of the spherical resonator, it perfectly describes the experimental results for the fundamental mode of the spherical resonator (as presented below).

Numerically, the resonant modes of the unloaded THz bubble resonator are studied using finite-difference time-domain (FDTD) simulations. The FDTD mode analysis is performed with the 'Harminv' package of the open source software MIT Electromagnetic Equation Propagation (MEEP) \cite{OskooiRo10,harminv}. For simplicity the simulations are implemented in cylindrical coordinates and possible deviations from a perfect sphere are not taken into account. The FDTD simulations predict the resonance frequencies of the measured THz bubble resonator and enable to identify the modes observed in the experiment. The simulations are based on the measured parameters of the THz bubble resonators.

\begin{figure}[t]
\centering
\fbox{\includegraphics[width=\linewidth]{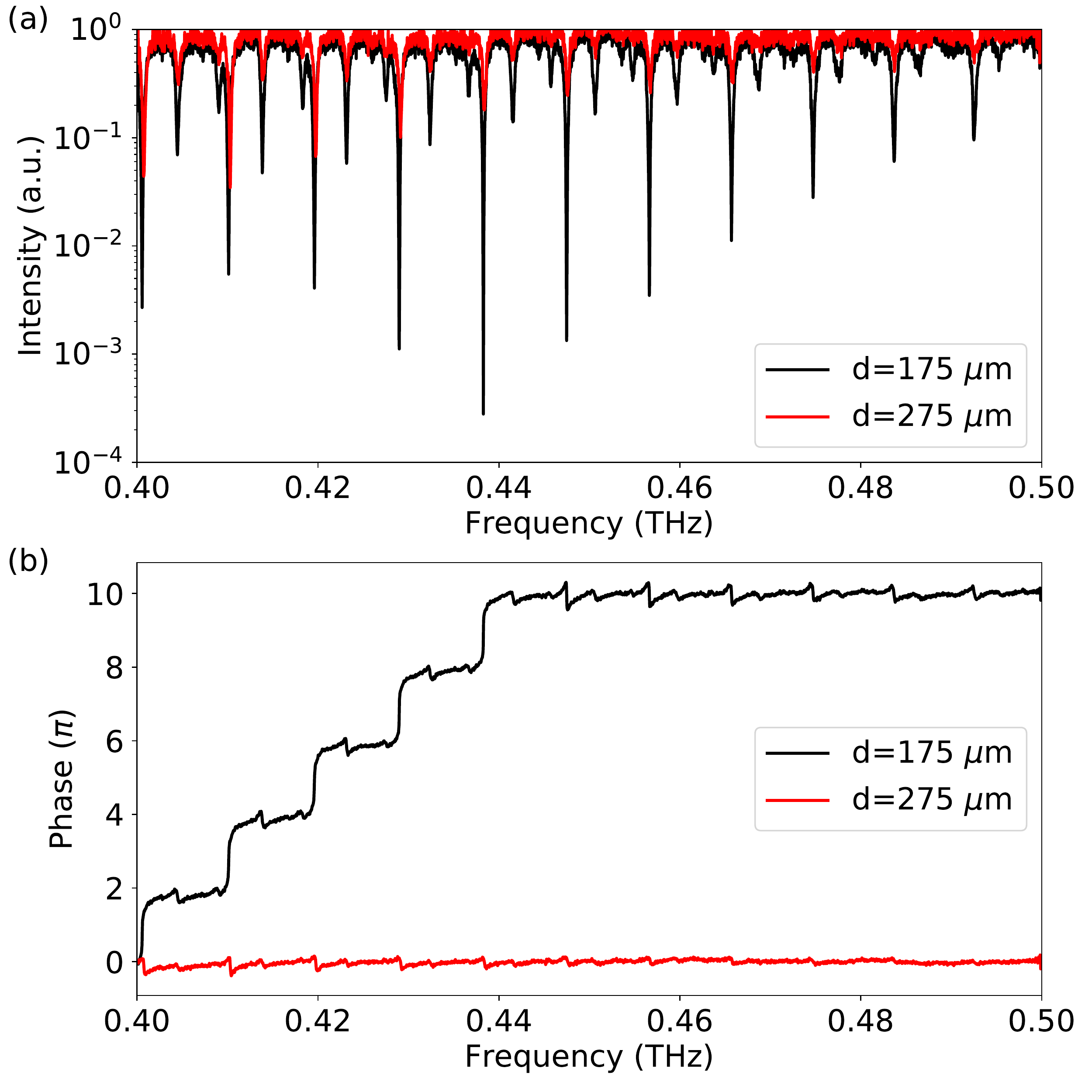}}
\caption{(a) Intensity of the measured transmission ratios in the frequency range from 0.4\,THz to 0.5\,THz on a logarithmic scale. The black solid line corresponds to a bus waveguide-resonator distance of about 175\,µm while the red solid line was recorded 100\,µm further away from the bus waveguide at about 275\,µm. The distances of 175\,µm and 275\,µm correspond to about ${\lambda}_{\textrm{quartz}}/1.9$ and ${\lambda}_{\textrm{quartz}}/1.2$ at 0.45\,THz, respectively. The black solid line is red shifted compared to the undercoupled case (red solid line) and (b) corresponding phase of the measured amplitudes shown above. Only the mode at 0.438\,THz is near critical coupling (steepest phase transition), while modes at lower and higher frequencies are overcoupled or undercoupled, respectively.} 
\label{fig:2}
\end{figure}

Fig. \ref{fig:2} (a) shows the intensity of the transmission ratios of the bus waveguide for two different bus waveguide-resonator distances over a wide frequency range from 0.4\,THz to 0.5\,THz. Please note the logarithmic scale. The black solid line corresponds to a bus waveguide-resonator distance of about 175\,µm (close to critical coupling), while the red solid line was measured at a distance of about 275\,µm (weak coupling). In both curves eleven strongly pronounced resonances are clearly visible. In particular, for the case of close to critical coupling (black solid line) the resonance at 0.438\,THz shows a drop in the intensity ratio by almost four orders of magnitude. The black solid line demonstrates a strong dependence of the coupling on the bus waveguide-resonator distance. The strength (i.e. drop in the transmission ratio) of the resonances decreases nearly symmetrically around the strongest coupled mode at 0.438\,THz. This observation can be qualitatively explained, considering the large 
wavelength change of 20$\%$ in 
the frequency range from 0.4\,THz (${\lambda}_{0}$ = 750\,µm) to 0.5\,THz (${\lambda}_{0}$ = 600\,µm). The latter highly affects the overlap of the evanescent fields of the bus waveguide and the WGMs. Furthermore, experiments have shown that critical coupling can be achieved for all modes using different bus waveguide - resonator distances, and therefore phase matching is not critical.

We also observe a small red shift of <\,0.2\,GHz for the resonance frequencies by approaching the THz bubble resonator to the bus waveguide. The latter can be qualitatively seen in fig. \ref{fig:2} (a) by comparing the resonance frequencies of the strong coupled case (black solid line) and the weakly coupled case (red solid line). The red shift is particularly pronounced for the resonances at lower frequencies (< 0.44\,THz) in fig. \ref{fig:2} (a). The red shift is caused by an increase of the effective refractive index in the evanescent field of the bubble resonator compared to the case without a bus waveguide (${n}_{silica} > {n}_{air}$) \cite{Preu:08}.

\begin{figure}[t]
\centering
\fbox{\includegraphics[width=\linewidth]{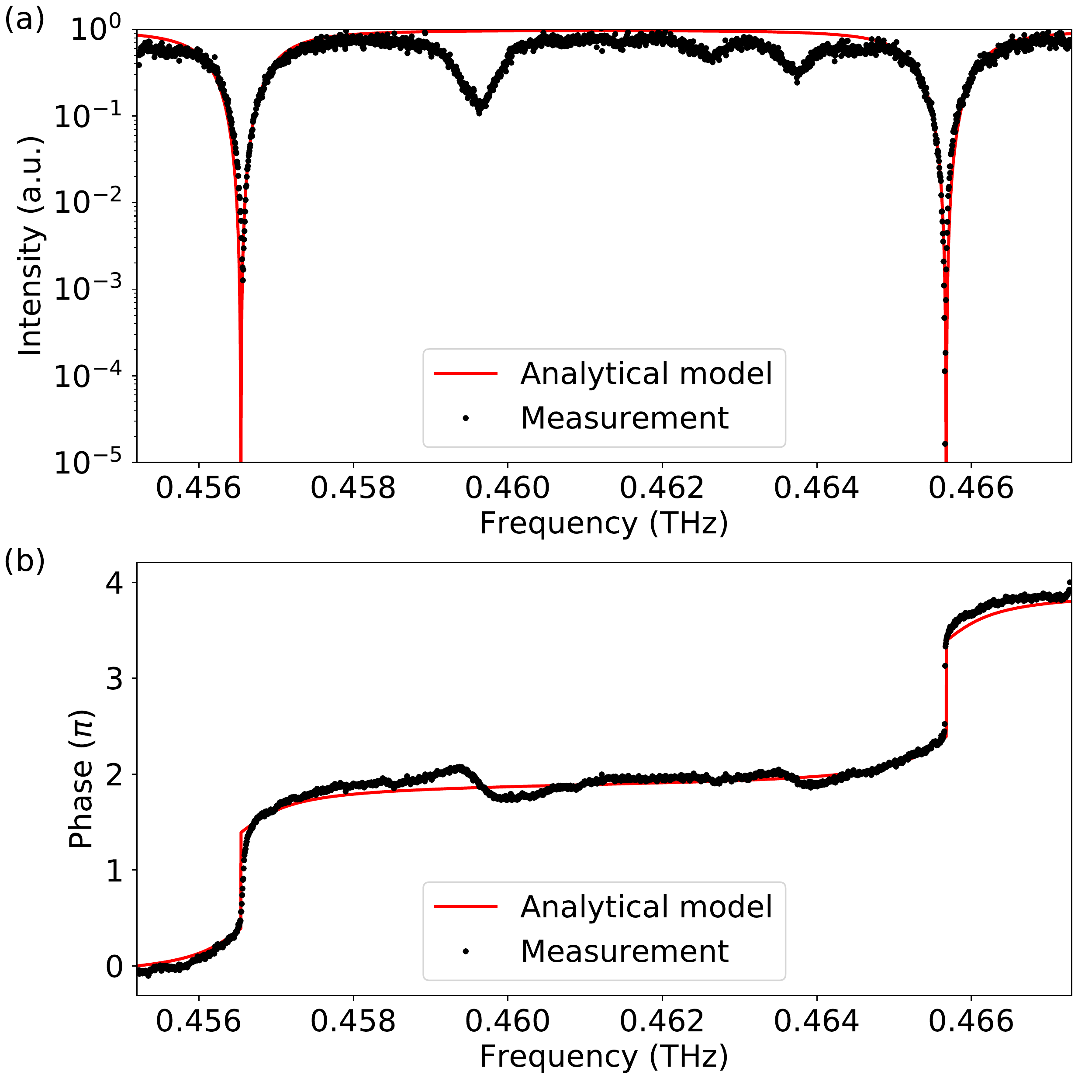}}
\caption{(a) Intensity and (b) phase of the measured transmission ratio (black dots) of the bus waveguide at critical coupling for the ${\textrm{TE}}_{40}$ mode (0.4657\,THz) in the frequency range from 0.455\,THz to 0.467\,THz. The bus waveguide-resonator distance is 160\,µm. The fit of the analytical model with $a=r=0.84$ is shown with the red solid line. An excellent agreement for both the intensity and phase is observed. In particular the step function like behavior in the phase profile for critical coupling is perfectly represented for the ${\textrm{TE}}_{40}$ mode. The finesse is 8.7 and the Q-factor of the ${\textrm{TE}}_{40}$ at 0.4657\,THz retrieved from the fit is 443.}
\label{fig:4}
\end{figure}

Each resonance has a specific fingerprint in the phase profile (shown in fig. \ref{fig:2} (b)) which allows conclusions about its coupling state - overcoupling, critical coupling and undercoupling. The states are well explained with the analytical expression for the effective phase shift induced by a ring resonator (see eq. \ref{eq:phase}). Overcoupled modes ($r < a$) can be identified by a slow ( i.e. over a wide frequency range of typically $>\,$60MHz) transition of 2$\pi$ in the phase profile, as observed for the modes at the lower end of the shown frequencies for the black solid curve (< 0.43\,THz). By approaching the condition for critical coupling ( i.e. coupling and resonator loss compensate: $a = r$), the phase shift occurs over a much narrower frequency range (as can be best seen by comparison of the two modes depicted in fig. \ref{fig:4} (b)). Finally, at critical coupling the phase profile also shows an overall 2$\pi$ shift but with a step function like behavior rather than a slow transition. The 
latter can be beautifully observed in fig. \ref{fig:4} (b) as discussed later in the text. Finally, undercoupling ($r > a$) can be identified by the phase profile seen for the higher frequencies above 0.44\,THz for the black solid curve and for all frequencies for the red solid curve ( large bus waveguide-resonator distance). The observed phase profile is in perfect agreement with the expectations. For a fixed distance of 175\,µm (black solid line) modes at low frequencies are overcoupled (the bus waveguide-resonator distance is too small for the wavelengths), while the modes at the higher frequencies are undercoupled (the bus waveguide-resonator distance is too large). In between those cases the mode at 0.438\,THz shows the steepest transition in the phase profile, a typical indication that the mode is very close to critical coupling. This phase information offers a complimentary information to the amplitude and shows the strong wavelength dependence of the coupling even for neighboring modes.

\begin{figure}[t]
\centering
\fbox{\includegraphics[width=\linewidth]{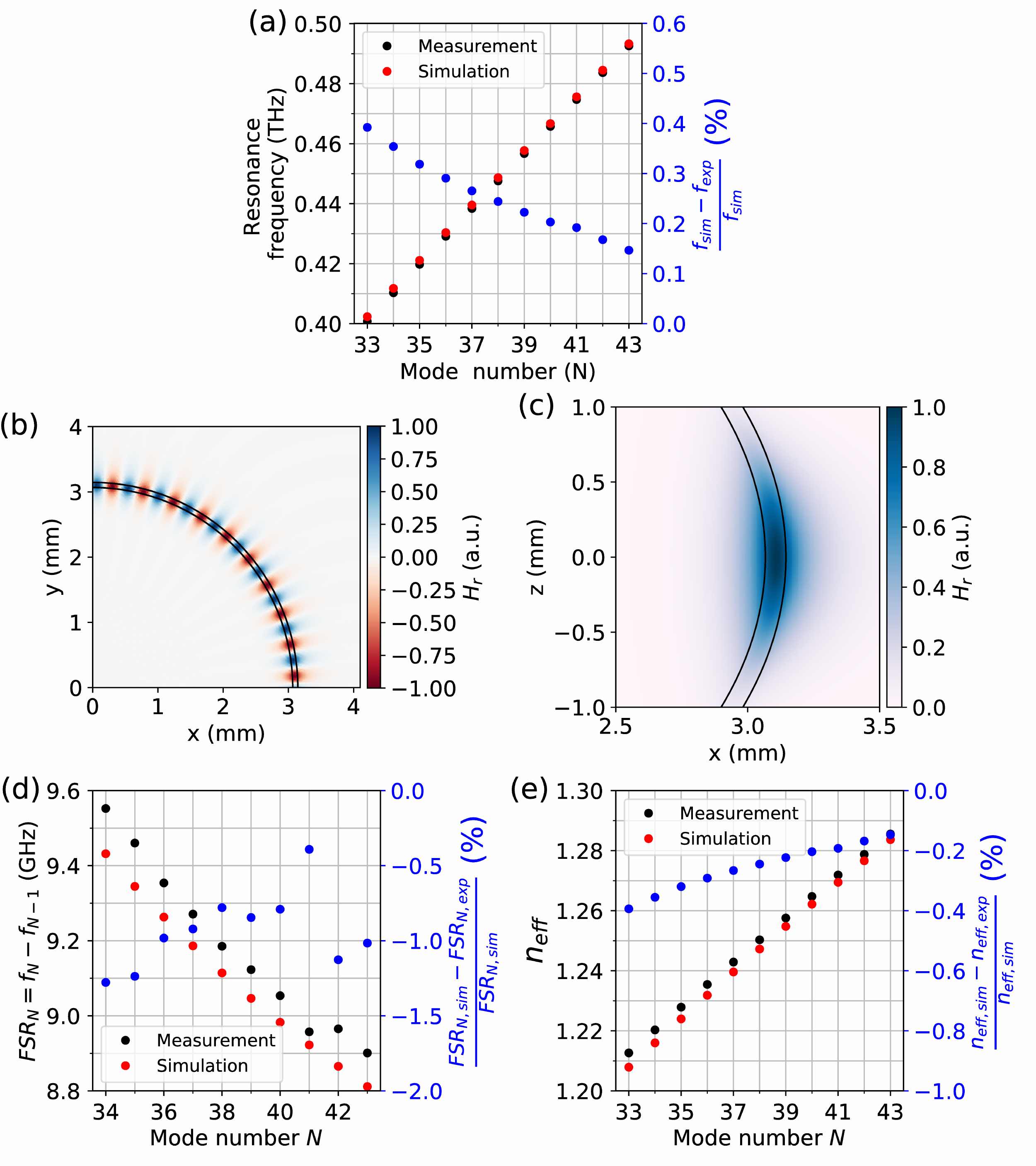}}
\caption{(a) Measured and simulated resonance frequencies extracted from the red solid line in fig. \ref{fig:2} (a) as a function of simulated mode number and (b), (c) visualize the normalized radial component of the magnetic field of the ${\textrm{TE}}_{40}$ mode in equatorial cross section and azimuthal cross section, respectively. Subfigure (d) depicts the measured and simulated ${FSR}_{N}={f}_{N} - {f}_{N-1}$ for the modes ranging from ${\textrm{TE}}_{33}$ to ${\textrm{TE}}_{43}$ and (e) shows the frequency dependent effective refractive index causing the highly changing FSR. The corresponding percentage differences between experiment and simulation are plotted in blue dots in subfigures (a), (d) and (e).}
\label{fig:3}
\end{figure}

Fig. \ref{fig:3} (a) depicts the measured and simulated resonance frequencies from the weakly coupled case (red line) presented in fig. \ref{fig:2} over the mode number retrieved from simulations. Comparison with simulations shows that the analyzed resonances are the fundamental ${\textrm{TE}}_{33}$ to ${\textrm{TE}}_{43}$ modes. Simulations also show that the sub-wavelength wall of the bubble does not support higher order radial modes in the analyzed frequency range from 0.4\,THz to 0.5\,THz. A thorough investigation of the higher order azimuthal modes present in the measurements (spectrally located between the fundamental modes but much weaker) is subject for further investigations. To visualize the very good agreement between the measured and predicted resonance frequencies of the fundamental modes, the relative frequency difference $({f}_{sim}-{f}_{exp})/{f}_{sim}$ is plotted in fig. \ref{fig:3} (a) with blue dots. A maximum deviation of <\,0.5\,$\%$ between experiment and simulation can be achieved by 
using an outside bubble diameter of 6.29\,mm and a 
wall thickness of 78\,µm as parameters. Fig. \ref{fig:3} (b) and (c) visualize the normalized radial component of the magnetic field of the analyzed ${\textrm{TE}}_{40}$ mode for an equatorial cross section and azimuthal cross section, respectively. The large extent of the evanescent field renders the THz bubble resonator an ideal candidate for a highly sensitive sensor. E.g., numerical results considering only changes inside the bubble indicate a very high resolution of better than $1 \times {10}^{-4}$ RIU. However, this is not even exploiting any of the strong material resonances present in the THz frequency range.   

Furthermore, comparing the frequency spacing between two adjacent modes of the weakly coupled case (red line in fig. \ref{fig:2}) reveals a constantly decreasing free spectral range (FSR) for higher frequencies. In the following the FSR between two adjacent modes is defined as ${FSR}_{N}={f}_{N} - {f}_{N-1}$, with $N$ being the mode number. The measured ${FSR}_{N}$ is plotted in fig. \ref{fig:3} (d) with black dots and demonstrates an average change of $\sim$\,70\,MHz per increment. This observation is well represented in the numerical data as can be seen by the percentage difference between experiment and simulation plotted with blue dots in fig. \ref{fig:3} (d). The variable FSR can be explained by a highly frequency dependent effective refractive index of each mode commonly referred to as waveguide dispersion. Fig. \ref{fig:3} (e) shows the measured (black dots) and simulated (red dots) effective refractive indices for the fundamental ${\textrm{TE}}_{33}$ to ${\textrm{TE}}_{43}$ modes (percentage 
difference is shown with blue dots). Experiment and simulation clearly show the trend of an increasing effective refractive index with increasing mode number. The latter can be qualitatively explained by the fact that higher mode numbers (smaller wavelength) are more strongly confined to the bubble wall, and therefore have an effective refractive index closer to the material refractive index of quartz glass (${n}_{\textrm{quartz}}=1.96$).

Due to the highly wavelength dependent coupling and the variable FSR, both effects which are not included in the applied analytical model \cite{silicon_microring}, we limit the following discussion to two resonances over a narrower frequency range. Fig. \ref{fig:4} (a) and (b) show the intensity and phase of the measured transmission ratio (black dots) with two resonances in the frequency range from 0.455\,THz to 0.467\,THz, respectively.  The modes are the fundamental ${\textrm{TE}}_{39}$ and ${\textrm{TE}}_{40}$. The bus waveguide-resonator distance is optimized to achieve critical coupling for the fundamental ${\textrm{TE}}_{40}$ mode. At a bus waveguide-resonator distance of 160\,µm a drop in the intensity ratio by up to $1.6\times {10}^{-5}$ for the ${\textrm{TE}}_{40}$ mode is observed. Furthermore, the step function like behavior in the measured phase (black dots in fig. \ref{fig:4} (b)) at the resonance frequency of the ${\textrm{TE}}_{40}$ mode confirms that the mode is indeed at critical coupling. 
For comparison, please note the much wider phase transition for the overcoupled ${\textrm{TE}}_{39}$ mode. For further analysis we fit the analytical model (see eq. \ref{eq:amplitude}) to the intensity ratio shown in fig. \ref{fig:4} (a). An excellent agreement is observed between the measurement and the analytical model, and also the phase profile of the ${\textrm{TE}}_{40 }$ mode is very well reproduced. The finesse obtained from the fit is 9 and the critically coupled ${\textrm{TE}}_{40}$ mode at 0.4657\,THz has a very high loaded Q-factor of 443.

\medskip

In conclusion, we successfully demonstrated WGMs in a dielectric THz  bubble resonator over a wide frequency range from 0.4\,THz to 0.5\,THz. We measured more than ten high Q-factor WGMs, identified as the fundamental ${\textrm{TE}}_{33}$ to ${\textrm{TE}}_{43}$ modes. For the first time, we were able to show very good agreement between the predicted phase profile of WGMs and experimental THz FDS measurements, especially showing the step function like behavior at critical coupling. We observed a red shift of the resonance frequencies due to the decreasing bus waveguide-resonator distance, and analyzed the frequency dependent FSR caused by waveguide dispersion. Both effects are experimentally accessible due to the small size of the bubble resonator compared to the THz wavelengths. Intriguingly, standard glass blowing techniques are sufficient to produce high Q-factor bubble resonators with sub-wavelength wall thicknesses. In particular, THz WGMs with a finesse of 9 and a very high Q-factor of >\,440 at 
critical coupling at 0.47\,THz are observed. Future work will concentrate on the design optimization of the dielectric WGM THz bubble resonator and its exploitation as sensor for minute changes in the concentration of gases and liquids.  



\end{document}